\documentclass[prl,twocolumn,showpacs,superscriptaddress]{revtex4}

\usepackage{amsmath}
\usepackage[thickspace,cdot]{SIunits}
\usepackage[dvips]{hyperref}
\usepackage[dvips]{graphicx}

\newcommand{\iA}{\ensuremath{\angstrom^{-1}}}
\renewcommand*{\vec}[1]{\ensuremath{\mathbf{#1}}}
\newcommand*{\vG}{\ensuremath{{\vec G}}}
\newcommand*{\vQ}{\ensuremath{{\vec G_0}}}
\newcommand*{\vO}{\ensuremath{{\vec 0}}}
\newcommand*{\vq}{\ensuremath{{\vec q}}}
\newcommand*{\vb}{\ensuremath{{\vec b}}}
\newcommand*{\veta}{\ensuremath{{\boldsymbol\eta}}}
\newcommand*{\vqred}{\ensuremath{\vq_{r}}}
\newcommand*{\pot}{\ensuremath{\varphi}}
\newcommand*{\eps}{\ensuremath{\epsilon}}
\newcommand*{\epsi}{\ensuremath{\eps^{-1}}}
\newcommand*{\w}{\ensuremath{\omega}}

\begin{document}

\preprint{submitted to Phys. Rev. Lett.}  
\title{Anomalous angular dependence of the dynamic structure factor near Bragg
reflections: Graphite}

\author{R.~Hambach}
\affiliation{Laboratoire des Solides Irradi{\'e}s, Ecole Polytechnique, CNRS, CEA/DSM, 91128 Palaiseau, France}
\affiliation{European Theoretical Spectroscopy Facility (ETSF)}
\affiliation{Institut f{\"u}r Festk{\"o}rpertheorie und -optik, Friedrich-Schiller-Universit{\"a}t Jena, 07743 Jena, Germany}

\author{C.~Giorgetti}
\affiliation{Laboratoire des Solides Irradi{\'e}s, Ecole Polytechnique, CNRS, CEA/DSM, 91128 Palaiseau, France}
\affiliation{European Theoretical Spectroscopy Facility (ETSF)}

\author{N.~Hiraoka}
\affiliation{National Synchrotron Radiation Research Center, Hsinchu 30076, Taiwan}

\author{Y.~Q. Cai}
\altaffiliation[current address: ]{National Synchrotron Light Source II, Brookhaven National Laboratory, Upton, NY 11973.}
\affiliation{National Synchrotron Radiation Research Center, Hsinchu 30076, Taiwan}

\author{F.~Sottile}
\affiliation{Laboratoire des Solides Irradi{\'e}s, Ecole Polytechnique, CNRS, CEA/DSM, 91128 Palaiseau, France}
\affiliation{European Theoretical Spectroscopy Facility (ETSF)}

\author{A.~G.~Marinopoulos}
\affiliation{CEMDRX, Department of Physics, University of Coimbra,
P-3004-516 Coimbra, Portugal}

\author {F.~Bechstedt}
\affiliation{Institut f{\"u}r Festk{\"o}rpertheorie und -optik, Friedrich-Schiller-Universit{\"a}t Jena, 07743 Jena, Germany}
\affiliation{European Theoretical Spectroscopy Facility (ETSF)}

\author{Lucia~Reining}
\affiliation{Laboratoire des Solides Irradi{\'e}s, Ecole Polytechnique, CNRS, CEA/DSM, 91128 Palaiseau, France}
\affiliation{European Theoretical Spectroscopy Facility (ETSF)}

%\date{submitted to Phys. Rev. Lett., \today}

%\bigskip

\begin{abstract}

  The electron energy--loss function of graphite is studied for momentum transfers   $\vq$ beyond the first Brillouin zone. We find that near Bragg   reflections the spectra can change drastically for very small variations in $\vq$. The effect is investigated by means of first principle calculations in the random   phase approximation and confirmed by inelastic x-ray   scattering  measurements of the dynamic structure factor $S(\vq,\w)$. We demonstrate that this effect is governed by crystal local field effects and the stacking of graphite. It is traced back to a strong coupling between excitations at small and large momentum transfers.

\end{abstract}


\maketitle

The momentum resolved and frequency dependent dynamic structure factor $S(\vq,\w)$ is a fundamental quantity in plasma physics, nuclear physics, particle physics and condensed matter physics. It is important for the understanding of many problems like e.g. electronic correlation, and it links the theory of many-body systems to scattering experiments like electron energy-loss spectroscopy (EELS) or inelastic x-ray scattering (IXS).

Most EELS measurements are restricted to moderate momentum transfer with $\vq$ typically being shorter than a reciprocal lattice vector due to multiple scattering. IXS, while being less appropriate in the range of very small \vq, does not suffer from this restriction \cite{PRB_Hiraoka_2005,PRL_Cai_2006}. Modern synchrotron radiation sources have therefore opened the way to study electronic excitations at large momentum transfer. New phenomena can be observed in this range such as a periodic plasmon dispersion in magnesium diboride MgB$_2$ \cite{PRL_Cai_2006} or plasmon-bands in Silicon \cite{PRL_Schuelke_1991}. Crystal local field effects (LFE) \cite{PR_Adler_1962,PR_Wiser_1963}, namely the fact that an external perturbation can induce a response on the length scale of the structure of the material, become increasingly important for large momentum transfer~\cite{PRB_Waidmann_2000}. They acquire particular importance for layered systems, such as graphite, or superlattices \cite{PRL_Botti_2002}. 
Electronic excitations in layered systems can reveal aspects that differ from the material in its ground state. Even for a system where the ground state charge density is mainly confined to the individual layers, excitations can induce Coulomb potentials that lead to significant interlayer interaction. One finds for example that plasmon frequencies are very sensitive to the interlayer distance in graphite~\cite{PRB_Marinopoulos_2004}.

\par
Graphite, with its well separated and polarizable graphene sheets, is a very good candidate for the exploration of LFE-induced phenomena. EELS and IXS measurements (see e.\,g.\  \cite{ZPA_Zeppenfeld_1968,ZPA_Zeppenfeld_1971,PRL_Marinopoulos_2002,PRB_Hiraoka_2005}) have determined the energy-loss spectra and plasmon dispersion for a wide range of momentum transfer \vq , and the continuous change of the spectrum with change in direction of \vq\ has been studied~\cite{PRL_Marinopoulos_2002}. Calculations of the energy-loss function that are based on the homogeneous electron gas (see e.\,g.\  \cite{PRB_Visscher_1971,PRB_Grecu_1973,AP_Fetter_1974}) or the tight-binding model (see e.\,g.\ \cite{SSC_Huang_1997,PRB_Lin_1997}) have been used extensively to study in-plane properties of graphite. \textit{Ab-initio} calculations based on Density-Functional Theory (DFT) in its time-dependent extension (TDDFT) \cite{PRL_Runge_1984}, either in the adiabatic local density approximation (TDLDA) or even in the Random Phase Approximation (RPA), reproduce experimental plasmon spectra with very good precision~\cite{PRL_Marinopoulos_2002,PRB_Marinopoulos_2004}. These calculations have been successfully used to describe the angular dependence of energy-loss spectra for relatively moderate momentum transfer $\vq$. The range of larger $\vq$, instead, is much less explored.

\par
In the present work we demonstrate that this range offers access to striking phenomena. In particular, our \textit{ab-initio} calculations and IXS measurements reveal an anomaly in the angular dependence of the dynamic structure factor: For a momentum transfer close to certain reciprocal lattice vectors, we observe drastic changes in the spectra upon small variations in \vq. This discontinuous behavior should have important implications for the interpretation of measurements close to Bragg reflections in any strongly inhomogeneous system.

% Methods
\par
We performed first principle RPA calculations of the momentum resolved and frequency dependent dynamic structure factor $S(\vq,\w)$, which is directly related to the energy-loss function. The electronic ground state of graphite (Bernal stacking) was calculated in DFT-LDA (local density approximation) with \textsc{Abinit}~\cite{CMS_Gonze_2002}, using a plane-wave basis set~\footnote{We used 6144 k-points and an energy cutoff of 28   Hartree.} and norm-conserving pseudopotentials~\cite{PRB_Troullier_1991}. The Kohn-Sham bandstructure was then used to compute the independent particle polarizability $\chi^0$ with the \textsc{dp}-code~\footnote{http://www.dp-code.org; V. Olevano, {\it et al.},   unpublished.}. As a next step we calculate the longitudinal dielectric function $\eps$ and its inverse $\epsi$, that links the total to the external potential in linear response, $\pot_\text{tot} = \epsi \pot_\text{ext}$.  In RPA $\eps = 1-v\chi^0$, where $v$ is the Coulomb interaction. In a periodic system $\eps$ is a {\it matrix} in reciprocal lattice vectors (\vG,\vG'), and a function of the reduced component \vqred\ of momentum transfer inside the first Brillouin zone and of frequency \w. For a given momentum transfer $\vq = \vqred + \vQ$ (where \vQ\ is a reciprocal lattice vector) the dynamic structure factor is then ($n_0$ denotes the average electron density)
\begin{align*}
%\label{eq:s}
 S(\vq ,\w) = -\frac{q^2}{4\pi^2
n_0}\operatorname{Im}\bigl[\epsi_{\vQ\vQ}({\vqred},\w)\bigr];
\end{align*}
only diagonal elements of \epsi\ are needed.  However, as $\eps$ is \textit{not} diagonal for an inhomogeneous material, its inversion will mix matrix elements. One can understand the physics of this mathematical fact by expanding $\epsi = (1-v\chi^0)^{-1}$:
\begin{multline*}
%\label{eq:expand}
 \epsi_{\vQ\vQ}(\vqred,\w) = 1 + v_{\vQ}\chi^0_{\vQ\vQ}(\vqred,\w) \\
  +\sum_{\vG}v_{\vQ}\chi^0_{\vQ\vG}(\vqred,\w)v_{\vG}\chi^0_{\vG\vQ}(\vqred,\w)+\dots
\end{multline*}
The first order term gives the response of the independent particles to the external potential.  The second term is the response to the Hartree potential that is induced by the first order response. This self-consistent process is then continued to infinite order. As $\chi_0$ is a matrix, an external potential with momentum $\vqred + \vQ$ can induce spatial charge fluctuations, whose momentum $\vqred + \vG$ differs by any reciprocal lattice vector, and to which the system will \textit{also} respond; these are the LFE.

\par
Most often LFE are studied where the external perturbation is of very long wavelength ($\vq \to 0$ in the optical case), and results are modified by contributions with \textit{shorter} wavelength. However, when $\vq =\vqred + \vQ$ is bigger than a reciprocal lattice vector one can also have contributions from \textit{larger} wavelength, i.\,e.\ smaller $|\vG|<|\vQ|$. For example, LFE can couple plasmons from the first Brillouin zone $\vqred$ with excitations at large momentum transfer $\vq=\vqred+\vQ$, leading to a Fano resonance in silicon \cite{PRL_Sturm_1992} and lithium \cite{EPJB_Hoeppner_1998} or a periodic plasmon dispersion for momentum transfer perpendicular to the planes in magnesium diboride MgB$_2$ \cite{PRL_Cai_2006}. We will show in the following that this situation is particularly striking for graphite, where it gives rise to an unexpected discontinuity of the dynamic structure factor.

\begin{figure}[t]
\includegraphics[clip]{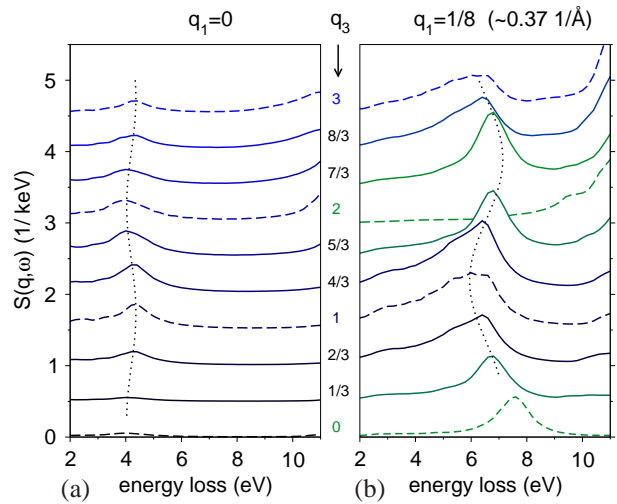}
\caption{(color online) Calculated dynamic structure factor $S(\vq,\w)$ of   graphite for different momentum transfers
$\vq=(q_1,0,q_3)$ (a) exactly perpendicular to the planes ($q_1=0$) and (b) with a   small in plane component ($q_1=\frac{1}{8}$). Dashed lines indicate spectra for integer $q_3$. All spectra have been shifted equidistantly in y-direction for better visibility. Dotted lines are guides to the eye.}
\end{figure}

\par
Figure~1(a) shows the calculated dynamic structure factor $S(\vq,\w)$ of graphite for a series of $\vq = (0,0,q_3)$ that are chosen exactly perpendicular to the planes [Note that $(q_1,q_2,q_3)= q_1\vb_1+q_2\vb_2+q_3\vb_3$ where the $\vb_n$ are primitive reciprocal lattice vectors; see Fig.~3(b)]. The structures that can be seen in this low-energy range are only due to the very weak interaction between graphene sheets; the intensity of the spectrum is therefore quite low and the dispersion small, contrary to the case of MgB$_2$ \cite{PRL_Cai_2006} where the coupling between layers is much stronger.

\par
Instead, Fig.~1(b) shows calculations for momentum transfers $\vq=(\frac{1}{8},0,q_3)$ where the in-plane component $ q_1 = \frac{1}{8}$ is kept fixed, and the perpendicular component $q_3$ is varied.  Although for large $q_3$ the direction of \vq\ deviates only slightly from the $c$-axis, we observe a striking difference compared to the on-axis results: First, the intensity is significantly increased for most of the spectra - when $q_3$ is zero, we simply see the in-plane $\pi$-plasmon. Second, there is a strong dispersion in the peak positions; the main peak shifts between $\unit{7.6}\electronvolt$ at $q_3=0$ and $\unit{6}\electronvolt$ at $q_3=1$. The most striking observation, however, is the disappearance of the peak at $q_3=2$: for that value, the dynamic structure factor abruptly becomes completely flat below $\unit{8}\electronvolt$. There is hence a {\it significant} change of the spectra for a {\it small} change of $\vq$. Even more important, the effect cannot be explained by the finite size of the in-plane component $q_1$: in the following we will show that these results remain valid for arbitrarily small $q_1$.

% Q=0
\par
The bottom panel of Fig.~2(a) shows $S(\vq,\w)$ calculated with and without LFE (solid and dashed lines, respectively) for momentum transfers $\vq=\veta$ that are vanishingly small ($|\veta|=\unit{5\cdot 10^{-5}}\iA$) and differ only in the angle $\theta$ to the c-axis. For in-plane momentum transfers ($\theta=\unit{90}\degree$), we find a pronounced $\pi$-plasmon peak at $\unit{7}\electronvolt$, whereas in perpendicular direction ($\theta=\unit{0}\degree$) a peak, although of much lower intensity, is found at $\unit{4}\electronvolt$. For intermediate directions of $\veta$ one finds a continuous behavior when the angle $\theta$ is changed, as one would expect. LFE contribute only marginally, i.e.\  the anisotropy simply stems from the anisotropy of the head element $\eps_{\vO\vO}(\veta,\w)$ of the dielectric matrix.  These results are consistent with earlier calculations on graphite \cite{PRB_Marinopoulos_2004}.

% Q=1
\par
Moving on to large momentum transfers, the next higher panel displays the results for $\vq=(0,0,1)+\veta$. A change in $\veta$ corresponds now to an infinitesimal change in $\vq$ [see Fig.~2(b)], and results both with and without LFE do not vary with $\veta$; therefore, only one result is shown. It should be noted however that LFE start to become significant, because of the increased momentum transfer.

% Q=2
\par
This picture changes completely when we reach momentum transfers $\vq=(0,0,2)+\veta$ near the second reciprocal lattice vector, shown in the top panel of Fig.~2(a): whereas the spectrum {\it without} LFE is completely flat and remains stable while $\veta$ is changed, we find that LFE lead to drastic modifications of the spectra for infinitesimal variations of the total momentum transfer \vq. A direct comparison between spectra for momentum transfers $\vq=(0,0,2)+\veta$ and the corresponding reduced momentum transfer $\veta$ shows that they are very similar besides a scaling factor, whenever \veta\ is not exactly in-plane (top vs.\ bottom panel); in other words, LFE lead to the reappearance of spectra of lower Brillouin zones \cite{PRL_Sturm_1992,PRL_Cai_2006}, and the dynamic structure factor $S$ near the reciprocal lattice vector $(0,0,2)$ is determined by the direction of the \textit{reduced} momentum transfer \vqred\ (here $\vqred=\veta$) and \textit{not} by the direction of \vq\ itself.
\begin{figure}[t]
\includegraphics{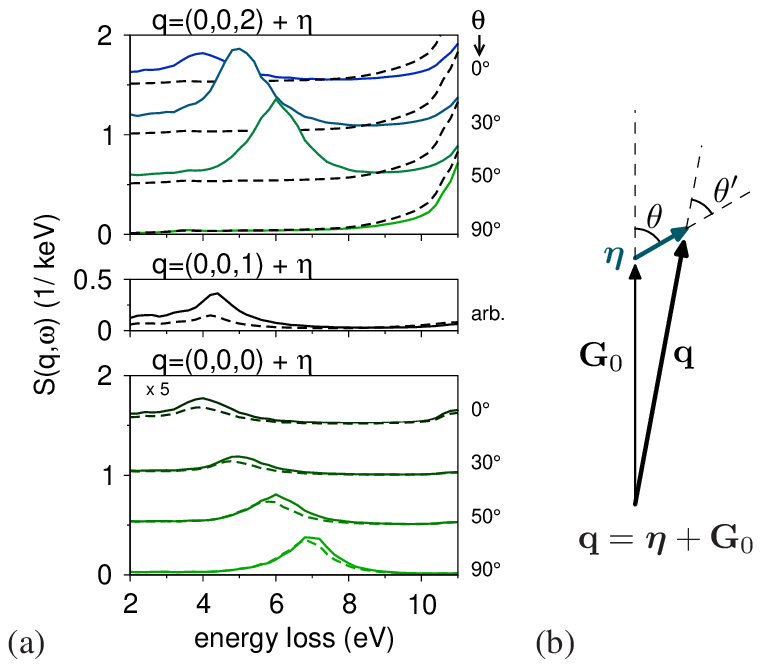}
\caption{(color online) (a) Dynamic structure factor $S(\veta+\vQ,\w)$   for small deviations $|\veta|=\unit{5\cdot 10^{-5}}\iA$ from   reciprocal lattice vectors $\vQ$ calculated with (solid) or without   (dashed) LFE. (Bottom) For $\vQ=(0,0,0)$, $S(\veta,\w)$ depends on   the direction of $\veta$. (Middle) For $\vQ=(0,0,1)$, arbitrary   $\veta$ lead to the same spectrum.  (Top) For $\vQ=(0,0,2)$,   $S(\veta+\vQ,\w)$ again changes with $\veta$, but only when LFE are   included. (b) Definition of the angles $\theta$ and $\theta'$.}
\end{figure}

\par
Graphite is a very convenient case for further analysis. Indeed, in the case of small \vqred\ and $\vQ=(0,0,2)$, we find a particularly strong coupling between \vqred\ and $\vqred + \vQ$ as the wing element $\eps_{\vO\vQ}(\vqred,\w)$ dominates by far all other off-diagonal elements $\eps_{\vG\vG'}(\vqred,\w)$. Consequently, neglecting the coupling to all other modes, the inversion of \eps\ reduces to the inversion of a simple $2\!\times\!2$ dielectric matrix~\footnote{Indeed, we have verified that results of calculations neglecting  all other off-diagonal elements are very close to the full   ones.} and one obtains (here, the $\w$ dependence is omitted):
\begin{align*}
%\label{slf}
\epsi_{\vQ\vQ}(\vqred) = \frac{1}{\eps_{\vQ\vQ}(\vqred)} +
\frac{\eps_{\vQ\vO}(\vqred)\eps_{\vO\vQ}(\vqred)}{[\eps_{\vQ\vQ}(\vqred)]^2}~\epsi_{\vO\vO}(\vqred).
\end{align*}
This result is similar to the two-plasmon-band approximation \cite{PRB_Sturm_1980}. The first term corresponds to the result without LFE. The second term leads to the reappearance of the spectrum $\epsi_{\vO\vO}(\vqred,\w)$. The latter is however weighted by the off-diagonal elements. Whenever the second term is strong, the spectrum for large $\vq=\vqred+\vQ$ will also depend on the one for the reduced component $\vqred$, and hence on its anisotropy.

\par
Still, it remains to be understood why this coupling to $\vq\to0$ does not show up {\it neither} for in-plane deviations $\vq=(\eta,0,2)$, {\it nor} around $\vQ=(0,0,1)$. To this end, we can make use of general properties of the dielectric function in semiconductors. In the limit of high frequencies, the coupling element $\eps_{\vO\vQ}(\vqred,\w)$ can be approximated as \cite{PSS_Bechstedt_1982}:
\begin{align*}
%\label{eq:wings}
 \epsilon_{\vO\vQ}(\vqred,\w) = \frac{4\pi}{\w^2}  
   ~\frac{\vqred\cdot(\vqred+\vQ)}{q_r^2}~n(-\vQ),
\end{align*}
where $n(-\vQ)$ denotes the Fourier coefficient of the electron density. A similar expression has been found by \citet{PRB_Sturm_1980} in the framework of a quasi-free electron gas. First, we see from this equation that the coupling between $\vqred+\vQ$ and $\vqred$ is proportional to the cosine of the angle $\theta'$ between the two corresponding directions [see Fig.~2(b)]. The prefactor $f=\eps_{\vQ\vO}\eps_{\vO\vQ}/\eps_{\vQ\vQ}^2\propto \cos^2 \theta'$ enforces the anisotropy of the spectra: in particular, for a small in-plane \vqred\ one has $\theta = 90\degree\approx\theta '$, which explains the absence of strong LFE in the spectrum for $\theta=90\degree$ [Fig.~2(a), top panel]. Second, the coupling vanishes whenever the Fourier component $n(-\vQ)$ of the electron density becomes zero. As this coefficient is proportional to the crystal structure factor, a wing element vanishes whenever the Bragg reflection of the corresponding reciprocal lattice vector is forbidden; for graphite in Bernal stacking, this is the case for all $\vQ=(0,0,2m+1)$ where $m$ is an integer.  LFE around $(0,0,1)$ stem hence only from a mixing with other {\it non-vanishing} ${\vG\not=\vO}$ beyond the $2\!\times\!2$ model. They do not introduce any significant dependence on the direction of $\vqred$ as $\vG+\vqred\approx\vG$. Instead, for $\vQ =(0,0, 2m)$ the two graphite planes A and B in the unit cell contribute in a constructive way (analogous to the constructive interference in the case of the Bragg reflection), which leads to the strong effect.

\par
With these explanations in mind, let us come back to the results shown in Fig.~1(b). Since $\vqred$ is still reasonably small, the above arguments hold. In particular, we can explain the drastic spectral changes near $(0,0,2)$ by the fact, that (i) the spectra from the first Brillouin zone reappears, which strongly depends on the direction of \vqred\ (angle $\theta$) due to the anisotropy of graphite, and (ii) the coupling and hence the strength of the recurring spectra is proportional to $|\vqred\cdot\vq|^2$ (angle $\theta'$).

\begin{figure}[t]
\includegraphics[clip]{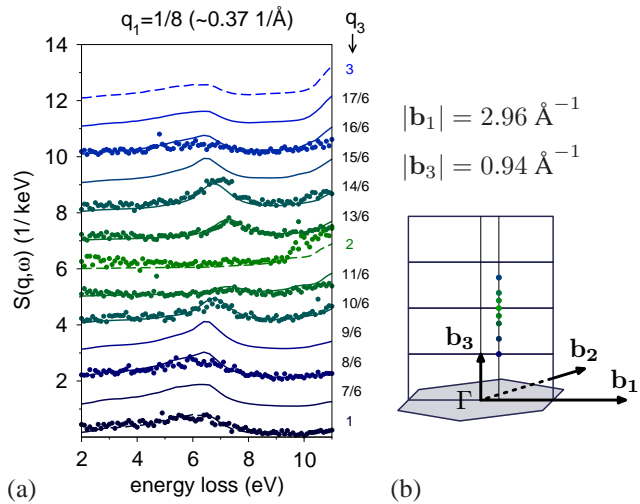}
\caption{(color online) (a) Comparison of the structure factor     
$S(\vq,\w)$ measured by IXS-experiments (dots) and calculated in     
RPA (lines; dashed for integer $q_3$) for     
$\vq=(\frac{1}{8},0,q_3)$. The elastic tail has been removed from     
the raw experimental data and a uniform scaling has been     
applied. (b) Measured $\vq$-points in reciprocal space.}
\end{figure}

\par
One may wonder, whether such a spectacular effect can actually be observed, or whether it is masked either by exchange-correlation effects that are neglected in RPA, or by experimental conditions like a strong elastic tail near the Bragg reflection, that prohibits a direct observation of the discontinuity at $(0,0,2)$. Therefore, we have conducted measurements of inelastic x-ray scattering for momentum transfers $\vq=(\frac{1}{8},0,\frac{n}{6})$ (for selected numbers $n$ between 6 and 16) covering the range between and above the Bragg reflections $(001)$ and $(002)$. The in-plane component $|q_1{\bf   b}_1|=\unit{0.37}\iA$ is still small enough to create the desired effect [see Fig.~1(b)], but large enough to avoid the Bragg reflection. The measurements were carried out at the Taiwan inelastic scattering beamline in \mbox{SPring-8} (BL12XU). The graphite sample was a plate having a surface of $2\times\unit{3}\milli\meter^2$ and a thickness of $\unit{100}\micro\meter$ \cite{PSJ_Suematsu_1972}. The x-ray Laue photograph showed very nice spots, indicating that the sample was not Highly Oriented Pyrolitic Graphite but a single crystal. The energy resolution was $\unit{140}\milli\electronvolt$. A Si 400 four-bounce monochromator and a Si 444 spherical crystal analyzer were used. The momentum resolutions were approximately $\unit{0.15}\iA$ along the horizontal and the vertical axes. In order to subtract the tails of the elastic lines for the spectra, glass was also measured as a reference. Fig.~3(a) shows the measured curves (dots), together with the corresponding calculated results (lines). The agreement is very good; in particular, the predicted peak shift is clearly seen in the measurements, as well as the abrupt change from a peaked spectrum for $\vq=(\frac{1}{8},0,\frac{13}{6})$ to a completely flat one at $\vq=(\frac{1}{8},0,2)$, and the difference between $q_3=\frac{13}{6} $ and $\frac{11}{6} $ due to the different angles $\theta'$ [see Fig.~2(b)]. Hence, our measurements give unambiguous support to the presented theoretical prediction.

% CONCLUSIONS
\par
In conclusion, our RPA calculations and IXS measurements revealed and explained a striking discontinuity in the dynamic structure factor $S(\vq,\w)$ of graphite at momentum transfers $\vqred + \vQ$ close to Bragg allowed reciprocal lattice vectors $\vQ$: infinitesimal changes in the momentum transfer induce strong changes in the resulting spectra. No discontinuity is observed when the crystal structure factor vanishes. It is hence a consequence of the Bernal stacking of the graphene layers that no changes occur at $\vQ=(0,0,1)$. Such a behavior of $S(\vq,\w)$ has important consequences for theory and measurements: First, anisotropic excitations from the first Brillouin zone might appear at large momentum transfers $\vq$ leading to an anomalous angular dependency of the spectra. Second, these recurring excitations belong to a direction \vqred\ \textit{different} from $\vq$. Third, from the experimental point of view, whenever measurements of the dynamic structure factor are performed close to an allowed Bragg reflection, the resulting spectra can be \textit{extremely   sensitive} to the chosen momentum transfer. We expect similar observations in other anisotropic crystals that show strong crystal local field effects, especially for layered or quasi 1D structures. 

{\bf Acknowledgements} This work was supported by the EU's 6th
Framework Programme through the NANOQUANTA Network of Excellence
(NMP4-CT-2004-500198) and by the ANR (project NT0S-3 43900).  The
experiment was carried out under an approval of SPring-8 and NSRRC
(Proposal No. C04B12XU-1510N). Y.~C.\ and N.~H.\ are grateful to Prof. Suematsu for providing us with the single crystal of graphite. R.~H.\ thanks the Dr.~Carl~Duisberg-Stiftung and C'Nano IdF (IF07-800/R).

% BIBLIOGRAPHY

\end{document}